\documentclass{article}
\usepackage{graphics}
\begin{document}

\title{Error Rate of the Kane Quantum Computer CNOT Gate in the Presence of Dephasing}

\author{Austin G. Fowler \and Cameron J. Wellard \and Lloyd C.L. Hollenberg}

\maketitle

\begin{abstract}
We study the error rate of CNOT operations in the Kane solid state
quantum computer architecture \cite{Kane98}. A spin Hamiltonian is
used to describe the system. Dephasing is included as exponential
decay of the off diagonal elements of the system's density matrix.
Using available spin echo decay data, the CNOT error rate is
estimated at approximately $10^{-3}$.
\end{abstract}

\section{Introduction}
\label{section:intro}

Existing classical computers manipulate bits that can be
exclusively $0$ or $1$. Quantum computers manipulate two level
quantum systems called qubits that can be arbitrary superpositions
of $|0\rangle$ and $|1\rangle$. While the idea of a quantum
computer was first suggested by Benioff and Feynman in the early
80s \cite{Beni80,Feyn82}, the first quantum algorithm that could
solve an interesting real world problem faster than it's classical
equivalent was published by Shor in $1994$ \cite{Shor94}. Shor's
algorithm factorizes integers with the number of steps growing
polynomially in the number of digits whereas the best known
classical algorithm grows exponentially. A significant milestone
in the quest to build a quantum computer was the construction of a
7 qubit liquid NMR implementation of Shor's algorithm designed to
factorize 15 \cite{Vand01}. Unfortunately, the liquid NMR approach
is not expected to work beyond a few tens of qubits \cite{Jone01}.

Many different technologies are being researched in the hope of
producing a scalable quantum computer. A sample of the diverse
proposals can be found in
\cite{Kane98,Enge01,Giov00,Hind01,Jame00,Mooi99}. In Kane's solid
state proposal \cite{Kane98,Kane00}, the nuclear spins of single
$^{31}$P dopant atoms in $^{28}$Si are used as qubits. This
approach aims to take maximum advantage of the industry expertise
acquired during the last 50 years of conventional semiconductor
electronics.

In this paper we study the error rate of CNOT operations in the
Kane quantum computer. Initial simulations were carried out without
dephasing to enable the pulse profiles of the controlling electrodes to be optimized
\cite{WelPhD}. The
lowest error rate achieved in the absence of dephasing was
$5\times 10^{-5}$. When a physically reasonable level of dephasing
was included in the simulation, the error rate increased to
approximately $10^{-3}$. While theoretical estimates of the error
rate required for fault tolerant computation are of order
$10^{-6}$ \cite{Knil97}, numerical simulations by Zalka suggest an
error rate of $10^{-3}$ or higher may be tolerable \cite{Zalk97}. Further work is
required to determine the maximum allowed error rate in the
Kane architecture.

This paper is organized as follows. In section \ref{section:kqc} the physical
architecture of the Kane quantum computer is described. In section
\ref{section:cnot} the process of performing a CNOT operation in the Kane
architecture is presented with emphasis on achieving the lowest possible error rate
in the absence of dephasing. Further details of this
process can be found in \cite{Goan00}. In section \ref{section:deph} our model of dephasing is
described and its effect on the error rate of the CNOT gate.
In section \ref{section:conc} we conclude with a discussion of the implications
of our estimate of the likely minimum error rate of the Kane CNOT gate.

\section{The Kane Quantum Computer}
\label{section:kqc}

The $^{31}$P in $^{28}$Si system is thought to be well suited to
use as a qubit due to it's long relaxation (T$_{1}$) and dephasing
(T$_{2}$) times. Both times only have meaning when the system is
in a steady magnetic field. Assuming the field is parallel with
the z-axis, the relaxation time refers to the time taken for $1/e$
of the spins in the sample to spontaneously flip whereas the
dephasing time refers to the time taken for the $x$ and $y$
components of a single spin to decay by a factor of $1/e$. In
natural silicon containing 4.7\% $^{29}$Si, relaxation times
T$_{1}$ in excess 1 hour have been observed for the donor electron
at T=1.25K and B$\sim$0.3T \cite{Fehe59}. The nuclear relaxation
time has been estimated at over 80 hours in similar conditions
\cite{Honi60}. The donor electron dephasing time T$_{2}$ in
enriched $^{28}$Si containing $(0.12\pm0.08)\%$ $^{29}$Si
\cite{Fehe58} has been measured at T=1.4K to be $\sim$0.5ms
\cite{Gord58}. At the time of writing, no experimental data
relating to the nuclear dephasing time has been obtained to the
authors' knowledge.

The phosphorous donor electrons are used primarily to mediate
interactions between neighboring nuclear qubits. As such, they are
polarized to remove their spin degree of freedom from the system.
This can be achieved by maintaining a steady B$_{z}$=2T at around
T=4K \cite{Goan00}. To take advantage of the long T$_{1}$ and
T$_{2}$ times discussed in the previous paragraph, the operating
temperature will more likely need to be $\sim$1K. Techniques for
relaxing the high field and low temperature requirements such as
spin refrigeration are under investigation \cite{Kane00}.

\begin{figure}[h]
\begin{center}
\resizebox{12cm}{!}{\includegraphics{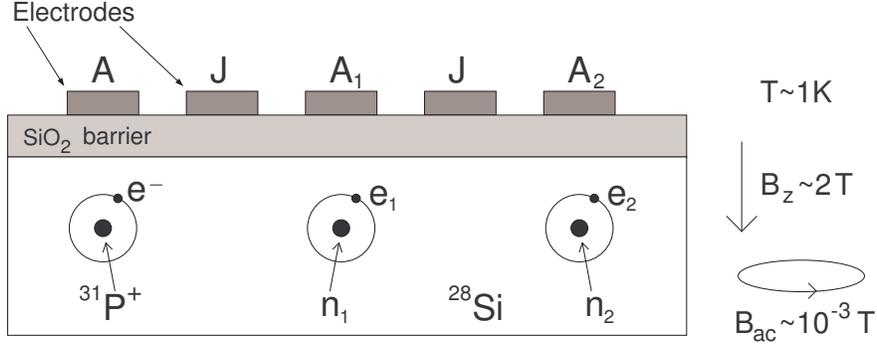}}
\end{center}
\caption{Schematic of the Kane architecture. The rightmost 2
qubits show the notation to be used when discussing the CNOT
operation.} \label{figure:3q}
\end{figure}

In the Kane architecture, qubits are arranged in a single line.
Control is achieved via electrodes above and between each qubit
and a global transverse oscillating field of magnitude $\sim$$10^{-3}$T (Fig.~\ref{figure:3q}).
To selectively manipulate a single qubit, the A-electrode above it
is biased. A positive/negative bias draws/drives the donor
electron away from the nucleus reducing the magnitude of the
interaction between the electron and nuclear spins. This in turn
reduces the energy difference between nuclear spin up
($|0\rangle$) and down ($|1\rangle$) allowing this transition to
be brought into resonance with a globally applied oscillating
magnetic field. Depending on the timing of the A-electrode bias,
the qubit can be rotated into an arbitrary superposition
$\alpha|0\rangle+\beta|1\rangle$. Clearly this scheme also allows
arbitrary combinations of individual qubits to be simultaneously
and independently manipulated.

Interactions between neighboring qubits are governed by the J-electrodes.
A positive bias encourages greater overlap of the
donor electron wave functions leading to indirect coupling of
their associated nuclei. In analogy to the single qubit case, this
allows two qubit transitions to be performed selectively between
arbitrary neighbors. A discussion of the electrode pulses required to
implement a CNOT gate is given in the next section.

\section{The CNOT Gate on a Kane QC}
\label{section:cnot}

Performing a CNOT operation on a Kane QC is an involved process
described in detail in \cite{Goan00}. Given the high field (2T)
and low temperature ($\sim$1K) operating conditions, we can model
the behavior of the system with a spin Hamiltonian. Only two
qubits are required to perform a CNOT operations so for the
remainder of the paper we will restrict our attention to a
computer with just two qubits. The basic notation is shown in
(Fig.~\ref{figure:3q}). Furthermore, let
\mbox{$\sigma^{z}_{n1}\equiv\sigma^{z}\otimes I\otimes I\otimes
I$}, \mbox{$\sigma^{z}_{e1}\equiv I\otimes\sigma^{z}\otimes
I\otimes I$}, \mbox{$\sigma^{z}_{n2}\equiv I\otimes
I\otimes\sigma^{z}\otimes I$} and \mbox{$\sigma^{z}_{e2}\equiv
I\otimes I\otimes I\otimes\sigma^{z}$} where $I$ is the $2\times
2$ identity matrix, $\sigma^{z}$ is the usual Pauli matrix and
$\otimes$ denotes the matrix outer product. With these definitions
the meaning of terms such as $\sigma^{y}_{n2}$ and
$\vec{\sigma}_{e1}$ should be self evident.

Let $g_{n}$ be the g-factor for the phosphorus nucleus, $\mu_{n}$
the nuclear magneton and $\mu_{B}$ the Bohr magneton. The
Hamiltonian can be broken into three parts
\begin{equation}
\label{H} H = H_{\rm Z}+H_{\rm int}(t)+H_{\rm ac}(t).
\end{equation}
The Zeeman interactions terms are contained in $H_{\rm Z}$
\begin{equation}
\label{HZ}
H_{\rm Z}=-g_{n}\mu_{n}B_{z}(\sigma^{z}_{n1}+\sigma^{z}_{n2})+\mu_{B}B_{z}(\sigma^{z}_{e1}+\sigma^{z}_{e2}).
\end{equation}
The contact hyperfine and exchange interaction terms, both of
which can be modified via the electrode potentials are
\begin{equation}
\label{Hint}
H_{\rm int}(t)=A_{1}(t)\vec{\sigma}_{n1}\cdot\vec{\sigma}_{e1}+A_{2}(t)\vec{\sigma}_{n2}\cdot\vec{\sigma}_{e2}+J(t)\vec{\sigma}_{e1}\cdot\vec{\sigma}_{e2},
\end{equation}
where $A_{i}(t) = 8\pi\mu_{B}g_{n}\mu_{n}|\Phi_{i}(0)|^{2}/3$, $|\Phi_{i}(0)|$
is the magnitude of the wavefunction of donor electron $i$ at
phosphorous nucleus $i$, and $J(t)$ depends on the overlap of the two donor
electron wave functions. The dependance
of these quantities on their associated electrode voltages is a subject of
ongoing research \cite{Koil02,Lari00,Vali99}. In this paper the hyperfine and
exchange interaction magnitudes $A_{i}$ and $J$ will frequently be discussed
as though directly manipulable.

The last
part of the Hamiltonian contains the coupling to the
global oscillating field $B_{\rm ac}$.
\begin{eqnarray}
\label{Hac}
H_{\rm ac}(t) & = & B_{\rm ac}(t)\cos(\omega t)[-g_{n}\mu_{n}(\sigma^{x}_{n1}+\sigma^{x}_{n2})
                                             +\mu_{B}(\sigma^{x}_{e1}+\sigma^{x}_{e2})] \nonumber \\
          & + & B_{\rm ac}(t)\sin(\omega t)[-g_{n}\mu_{n}(\sigma^{y}_{n1}+\sigma^{y}_{n2})
                                             +\mu_{B}(\sigma^{y}_{e1}+\sigma^{y}_{e2})].
\end{eqnarray}
Using the above definitions, only the
quantities $A_{1}$, $J$ and $B_{\rm ac}$ need to be manipulated to
perform a CNOT operation.

For clarity assume the computer is initially in one of the states
$|00\rangle$, $|01\rangle$, $|10\rangle$ or $|11\rangle$ and that
we wish to perform a CNOT operation with qubit 1 as the control.
Step one is to break the degeneracy of the two qubits' energy
levels to allow the control and target qubits to be distinguished.
To make qubit 1 the control the value of $A_{1}$ is increased
(qubit 1 will be assumed to be the control qubit for the remainder
of the paper).

Step two is to gradually apply a positive potential to the J electrode
in order to force greater overlap of the donor electron wave
functions and hence greater (indirect) coupling of the underlying
nuclear qubits. The rate of this change is limited so as to be
adiabatic --- qubits initially in energy eigenstates remain in
energy eigenstates throughout this step.

Let $|$symm$\rangle$ and $|$anti$\rangle$ denote the standard
symmetric and antisymmetric superpositions of $|10\rangle$ and
$|01\rangle$. Step three is to adiabatically reduce the $A_{1}$
coupling back to its initial value once more. During this step,
anti-level-crossing behavior changes the input states as
$|10\rangle\rightarrow|$symm$\rangle$ and
$|01\rangle\rightarrow|$anti$\rangle$.

\begin{figure}[p]
\begin{center}
\resizebox{12cm}{!}{\includegraphics{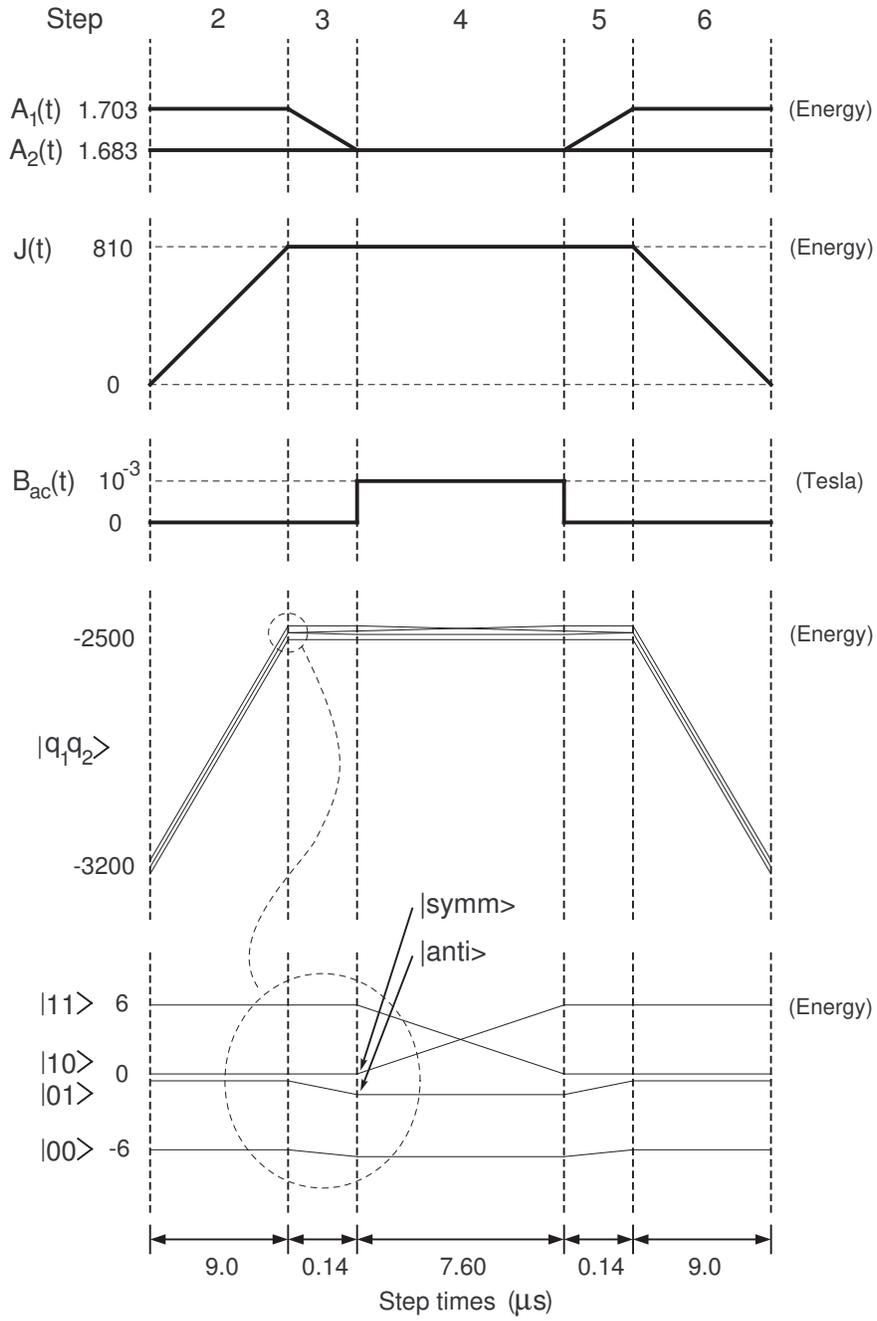}}
\end{center}
\caption{Gate profiles and state energies during a CNOT operation
in units of $g_{n}\mu_{n}B_{z}=7.1\times10^{-5}$meV.}
\label{figure:pulses}
\end{figure}

Step four is the application of an oscillating field $B_{\rm ac}$
resonant with the $|$symm$\rangle\leftrightarrow|11\rangle$
transition. This oscillating field is maintained until these two
states have been interchanged. Steps five to seven are the time
reverse of steps one to three. The process is shown schematically
in (Fig.~\ref{figure:pulses}). Note that steps 1 and 7 (the
increasing and decreasing of $A_{1}$) have been omitted as the
only limit to their speed is that they be done in a time much
greater than $\hbar/0.01$eV$\sim0.1$ps where 0.01eV is the orbital
excitation energy of the donor electron.

In general the fidelity of the adiabatic steps in the procedure
can be increased arbitrarily by making them indefinitely long. In
reality, of course, this is not practical and the procedure must
be implemented on a time-scale that is short compared to the
systems dephasing time. For a given $H(t)$, the degree to which
the evolution deviates from perfect adiabaticity can quantified by
the measure \cite{Bran89}
\begin{equation}
\Theta(t) \equiv {\rm Max_{a\neq b}} \left[ {\hbar | \langle
\psi_{\rm a}(t)| {\partial \over \partial t}(H(t)) | \psi_{\rm
b}(t)\rangle | \over (\langle \psi_{\rm a}(t)|H(t)| \psi_{\rm
a}(t)\rangle - \langle \psi_{\rm b} (t) |H(t)| \psi_{\rm
b}(t)\rangle )^2} \right]. \label{equation:adiabaticcond}
\end{equation}
It is desired that $\Theta(t)\ll 1$. Here the states ${|\psi_a(t)\rangle}$ are the set of eigenstates
of the Hamiltonian at time $t$. It is possible to reduce $\Theta(t)$
without increasing the duration of the step by optimizing the profiles of the
time dependent parameters in the Hamiltonian. In the case of
the Kane architecture this means optimizing the shape of the
evolution of $A_{1}(t)$ and $J(t)$.

\begin{figure}[h]
\begin{center}
\resizebox{8cm}{!}{\includegraphics{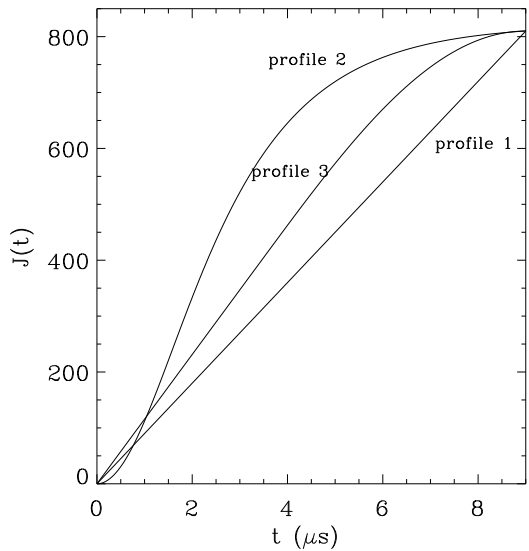}}
\end{center}
\caption{Possible forms of the J(t) profile for step 2 of the
adiabatic CNOT gate. J(t) is in units of
$g_{n}\mu_{n}B_{z}=7.1\times10^{-5}$meV.}
\label{figure:Adia J}
\end{figure}
\begin{figure}[h]
\begin{center}
\resizebox{8cm}{!}{\includegraphics{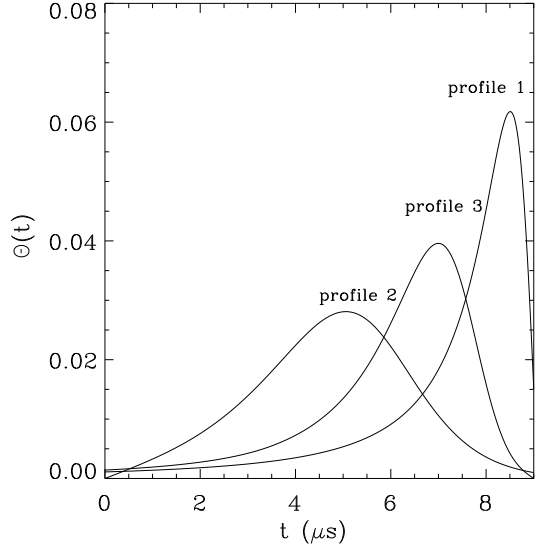}}
\end{center}
\caption{The adiabatic measure $\Theta(t)$ for each $J(t)$ profile.}
\label{figure:Adia Theta}
\end{figure}
Various profiles for the adiabatic steps in the CNOT procedure
have been investigated in \cite{WelPhD}. In (Fig.~\ref{figure:Adia J}), we
have plotted three possible $J(t)$ profiles for
step 2 of the CNOT gate. The function $\Theta(t)$ for each profile is
shown in (Fig.~\ref{figure:Adia Theta}).
Profile 1 is a simple linear pulse, profile 2 can be seen to be
the best of the three and is described by $J(t) = 810\alpha (1 -
{\rm sech}(5 t /\tau ))$ where $\tau=9\mu s$ is the duration of the
pulse and $\alpha = 1.0366$ is a factor introduced to ensure that
$J(\tau) = 810$. The third profile
\begin{equation}
J(t)=\left\{
\begin{array}{ll}
\frac{J_{max}}{2}\frac{t(1+\pi /2)}{T}, & 0<t<T/(1+\pi /2)  \\
\frac{J_{max}}{2} \left[ 1+\sin \left( \frac{\pi}{2}\frac{t-T/(1+\pi /2)}{T/(1+2/\pi )} \right)  \right], & T/(1+\pi /2)<t<T
\end{array}
\right.
\label{equation:linsin}
\end{equation}
although not quite as efficient as profile 2, is a composite
linear-sinusoidal profile that was used in the calculations
presented in this paper due to numerical difficulties in solving
the Schr\"odinger equation for profile 2. The advantage of the
second two profiles over the linear one is that they flatten out
as $J$ approaches $810$. At $J=816.65$, the system undergoes a
level crossing. To maintain adiabatic evolution, $J(t)$ needs
to change more slowly near this value. Note that the reason it is
desirable to make $J(t)$ so large is to ensure that there is a
large energy difference between $|$symm$\rangle$ and
$|$anti$\rangle$ during step 4 (the application of $B_{\rm ac}$).
This difference is given by
\begin{equation}
\delta E = 2 A^2({1 \over \mu_B B_z + g_n \mu_n B_z} - {1 \over \mu_B B_z + g_n \mu_n B_z-2J}).
\end{equation}
Without a large energy difference, the oscillating field
$B_{\rm ac}$ which is set to resonate with the transition
$|$symm$\rangle \leftrightarrow |11\rangle$ will also be very
close to resonant with $|$anti$\rangle \leftrightarrow |11\rangle$
causing a large error during the operation of the CNOT gate.
While it is desirable to make $J(t)$ large, it must be kept
comfortably below $816.65$ as near this level crossing the time
required to adiabatically increase
$J(t)$ increases significantly.

Step 3 (the decreasing of $A_{1}$) could be performed without degrading
the overall fidelity of the gate in a time of less than a
micro-second with a linear pulse profile.

The above steps were simulated using an adaptive Runge-Kutta
routine to solve the density matrix form of the Schr\"odinger
equation in the computational basis
$|n_{1}e_{1}n_{2}e_{2}\rangle$.
\begin{equation}
\label{rho dot}
\dot{\rho}(t)=\frac{1}{i\hbar}[H(t),\rho(t)].
\end{equation}
The times used for each stage are as follows

\begin{center}
\begin{tabular}{|c|c|} \hline
\emph{stage} & \emph{duration ($\mu s$)} \\ \hline
2 & 9.0000 \\
3 & 0.1400 \\
4 & 7.5989 \\
5 & 9.0000 \\
6 & 0.1400 \\ \hline
\end{tabular}
\end{center}

Note that the precision of the duration of stage 4
is required as the oscillating field $B_{\rm ac}$ induces the states
$|11\rangle$ and $|$symm$\rangle$ to swap smoothly back and forth.
The duration 7.5989$\mu s$ is the time required for one swap.
The other step times were obtained by first setting them to arbitrary values
($\sim$5$\mu s$) and increasing them until the gate fidelity ceased to increase.
The step times were then decreased one by one until the fidelity started
to decrease. As such, the above times are the minimum time in which the
maximum fidelity can be achieved. This maximum fidelity was found to be $5\times 10^{-5}$
for all computation basis states.

\section{Intrinsic dephasing and fidelity}
\label{section:deph}

In this paper, dephasing is modelled as exponential decay of the
off diagonal components of the density matrix. While a large
variety of dephasing models exist
\cite{Baue02,Burk99,Kimm02,Mozy01,Sous02,Thor01}, this
approximation is consistent with the first order nature of the
spin Hamiltonian. The donor electrons and phosphorous nuclei are
assumed to dephase at independent rates. With the inclusion of
dephasing terms (Eq.~\ref{rho dot}) becomes
\begin{eqnarray}
\label{rho dot dephase}
\dot{\rho} & = & \frac{1}{i\hbar}[H,\rho] \nonumber \\
           &   & -\Gamma_{e}[\sigma^{z}_{e_{1}},[\sigma^{z}_{e_{1}},\rho]]-\Gamma_{e}[\sigma^{z}_{e_{2}},[\sigma^{z}_{e_{2}},\rho]] \nonumber \\
           &   & -\Gamma_{n}[\sigma^{z}_{n_{1}},[\sigma^{z}_{n_{1}},\rho]]-\Gamma_{n}[\sigma^{z}_{n_{2}},[\sigma^{z}_{e_{2}},\rho]].
\end{eqnarray}
To understand the effect of each double commutator, it is instructive to
consider the following simple mathematical example :
\begin{eqnarray}
\label{simp dephase}
\dot{M} & = & -\Gamma[\sigma^{z},[\sigma^{z},M]] \nonumber \\
\left( \begin{array}{cc}
\dot{m}_{11} & \dot{m}_{12}  \\
\dot{m}_{21} & \dot{m}_{22}
\end{array} \right) & = &
\left( \begin{array}{cc}
0 & -4\Gamma m_{12} \\
-4\Gamma m_{21} & 0
\end{array} \right) \nonumber \\
\left( \begin{array}{cc}
m_{11}(t) & m_{12}(t)  \\
m_{21}(t) & m_{22}(t)
\end{array} \right) & = &
\left( \begin{array}{cc}
m_{11}(0) & m_{12}(0)e^{-4\Gamma t} \\
m_{21}(0)e^{-4\Gamma t} & m_{22}(0)
\end{array} \right).
\end{eqnarray}
Thus each double commutator in (Eq.~\ref{rho dot dephase})
exponentially decays it's associated off diagonal elements with a
characteristic time $\tau_{e}=1/4\Gamma_{e}$ or
$\tau_{n}=1/4\Gamma_{n}$.

\begin{figure}[p]
\begin{center}
\resizebox{12cm}{!}{\includegraphics{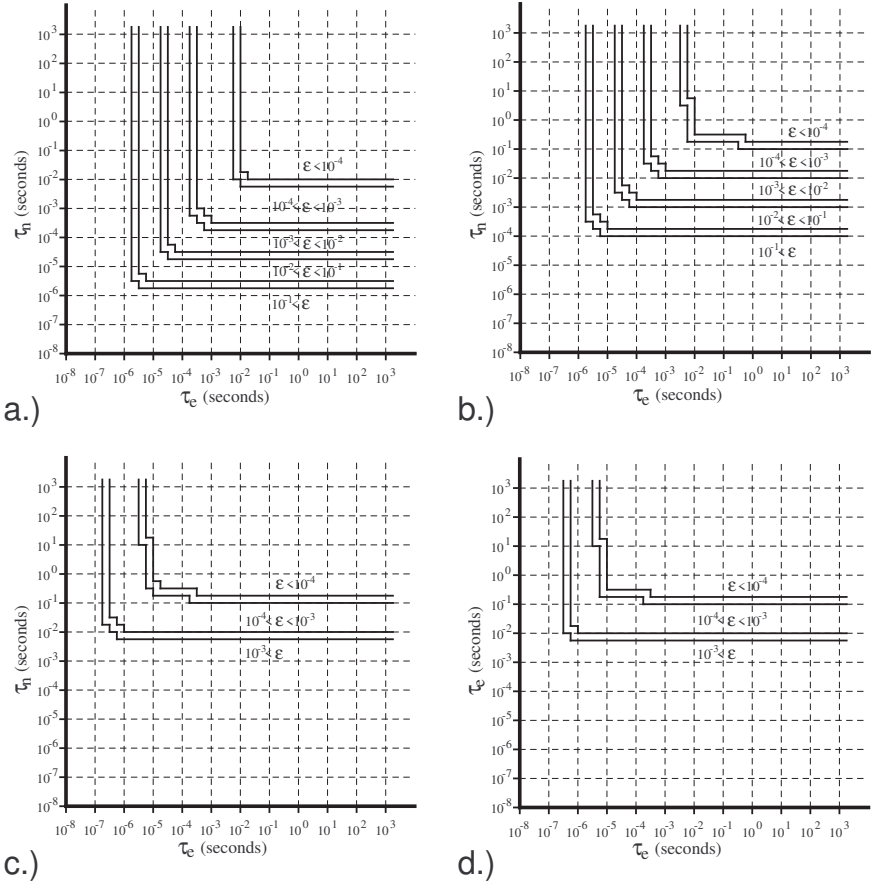}}
\end{center}
\caption{Probability of error during a CNOT operation as a function
of $\tau_{e}$ and $\tau_{n}$ for input state a.) $|00\rangle$, b.)
$|01\rangle$, c.) $|10\rangle$ and d.) $|11\rangle$. The first
qubit is the control.}
\label{figure:each}
\end{figure}

\begin{figure}[h]
\begin{center}
\resizebox{10cm}{!}{\includegraphics{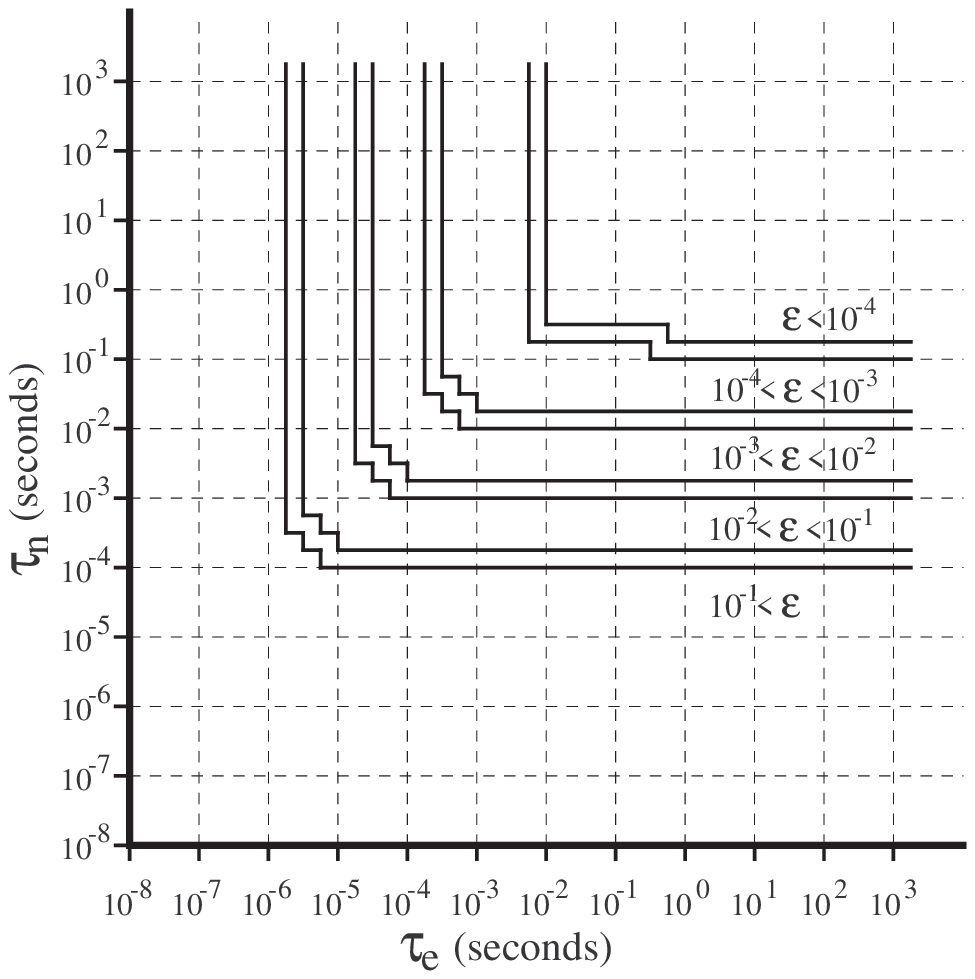}}
\end{center}
\caption{The worst case probability of error during a CNOT
operation as a function of $\tau_{e}$ and $\tau_{n}$ for all input
states.}
\label{figure:all}
\end{figure}

For each initial state $|00\rangle$, $|01\rangle$, $|10\rangle$
and $|11\rangle$ (Eq.~\ref{rho dot dephase}) was solved for a
range of values of $\tau_{e}$ and $\tau_{n}$ using the pulse
profiles described in section \ref{section:cnot} allowing a
contour plot of the gate error versus $\tau_{e}$ and $\tau_{n}$ to
be constructed (Fig.~\ref{figure:each}). Note that each contour is
a double line as each run of the simulation required considerable
computational time and the data available does not allow finer
delineation of exactly where each contour is. The worst case error
of all input states as a function of $\tau_{e}$ and $\tau_{n}$ is
shown in (Fig.~\ref{figure:all}).


\section{Conclusion}
\label{section:conc}

At the time of writing, to the authors' knowledge the only
experimental measurement of the $^{31}$P in $^{28}$Si dephasing
times in is \cite{Gord58} in which the donor electron dephasing
time T$_{2}$ was measured at T=1.4K, B$_{z}$=0.3T to be
$\sim$0.5ms. The $^{28}$Si sample contained $(0.12\pm0.08)\%$
$^{29}$Si \cite{Fehe58}. Note that a dopant concentration of
$\sim$$10^{16}$cm$^{-3}$ was used implying a donor separation of
$\sim$50nm. If this T$_{2}$ is used for $\tau_{e}$ and if
$\tau_{n}$ is assumed to be a couple of orders of magnitude larger
as in the case of the relaxation times, then from
(Fig.~\ref{figure:all}) the overall error probability would be
just under $10^{-3}$

Theoretical calculation of $\tau_{e}$ and $\tau_{n}$ has been
performed in \cite{Sous02} for a 2D array of P donors spaced 10nm
apart in pure $^{28}$Si yielding $\tau_{e}=2{\rm \mu s}$ and
$\tau_{n}=10$s. Such a short $\tau_{e}$ would imply an
unacceptable error probability of about 10\%. However, this same
paper also contains similar calculations for natural silicon
(4.7\% $^{29}$Si) with $\tau_{e}$ quoted as 200${\rm \mu s}$ which
leads to an overall error probability just over $10^{-3}$. The
suppression of decoherence in this case arises from line
broadening due to the presence of $^{29}$Si nuclei. In the case of
the Kane quantum computer, similar suppression can be achieved by
biasing the A-electrodes such that nearby qubits have different
spin-flip energies. Further investigation of this point is
required.

Though $10^{-3}$ is a large error probability, numerical
simulations by Zalka suggest this may be tolerable \cite{Zalk97}.
Work is in progress on simulations to determine an acceptable
error rate for the Kane architecture.

\section{Acknowledgements}
\label{section:ack}

The authors thank W. Haig (High Performance Computing System
Support Group, Department Of Defence), as well as the Victorian
Partnership for Advanced Computing for computational support. This
work was funded by the Australian Research Council.


\begin{thebibliography}{99}

\bibitem{Kane98} B. Kane, {\it Nature} {\bf 393}, 133(1998)
\bibitem{Beni80} P. Benioff, {\it Journal of Statistical Physics} {\bf 22}, 563(1980)
\bibitem{Feyn82} R. Feynman, {\it International Journal of Theoretical Physics} {\bf 21}, 467(1982)
\bibitem{Shor94} P. Shor, {\it Algorithmic Number Theory. First International Symposium, ANTS-I. Procedings}, 289(1994)
\bibitem{Vand01} L. Vandersypen {\it et. al}, {\it Nature} {\bf 414}, 883(2001)
\bibitem{Jone01} J. Jones, {\it Progress in Nuclear Magnetic Resonance Spectroscopy} {\bf 38}, 325(2001)
\bibitem{Enge01} H-A. Engel, P. Recher and D. Loss, {\it Solid State Communications} {\bf 119}, 229(2001)
\bibitem{Giov00} V. Giovannetti, P. Tombesi and D. Vitali, {\it Journal of Modern Optics} {\bf 21}, 467(1982)
\bibitem{Hind01} E. Hinds, {\it Physics World}, 39(2001)
\bibitem{Jame00} D. James, {\it Fortschritte der Physik} {\bf 48}, 823(2000)
\bibitem{Mooi99} J. Mooij, {\it Microelectronic Engineering} {\bf 47}, 3(1999)
\bibitem{Kane00} B. Kane, {\it Forschritte der Physik} {\bf 48}, 1023(2000)
\bibitem{WelPhD} C. Wellard, {\it PhD Thesis, The University of Melbourne}, (2001)
\bibitem{Knil97} E. Knill, R. Laflamme and W. Zurek, {\it quant-ph/9702058}, (1997)
\bibitem{Zalk97} C. Zalka, {\it quant-ph/9505011}, (1997)
\bibitem{Goan00} H.S. Goan and G.J. Milburn, {\it unpublished internal SRC report}, (2001)
\bibitem{Fehe59} G. Feher and E. Gere, {\it Physical Review} {\bf 114}, 1245(1959)
\bibitem{Honi60} A. Honig and E. Stupp, {\it Physical Review} {\bf 117}, 69(1960)
\bibitem{Fehe58} G. Feher and others {\it Physical Review} {\bf 109}, 221(1958)
\bibitem{Gord58} J.P. Gordon and K.D. Bowers, {\it Physical Review Letters} {\bf 10}, 368(1958)
\bibitem{Koil02} B. Koiler, X. Hu and S. Das Sarma, {\it cond-mat/0106259}, (2002)
\bibitem{Lari00} A. Larionov, L. Fedichken, A. Kokin and K. Valiev, {\it Nanotechnology} {\bf 11}, 392(2000)
\bibitem{Vali99} K. Valiev and A. Kokin, {\it quant-ph/9909008}, (1999)
\bibitem{Bran89} B.H. Bransden and C.J. Joachain, {\it Introduction to Quantum Mechanics}, (1989)
\bibitem{Baue02} W.Bauer and W. Nadler, {\it cond-mat/0204325}, (2002)
\bibitem{Burk99} G. Burkard and D. Loss, {\it Physical Review B} {\bf 59}, 2070(1999)
\bibitem{Kimm02} K. Kim and H. Kwon, {\it Physical Review A} {\bf 65}, 1(2002)
\bibitem{Mozy01} D. Mozyrsky, S. Kogan and G. Berman, {\it cond-mat/0112135}, (2001)
\bibitem{Sous02} R. de Sousa and S. Das Sarma {\it cond-mat/0203101}, (2002)
\bibitem{Thor01} M. Thorwart, {\it Physical Review A} {\bf 65}, 1(2001)
\bibitem{Gott99} D. Gottesman, {\it quant-ph/9903099}, (1999)

\end{thebibliography}
\end{document}